\documentclass[aip,apl,twocolumn,groupedaddress,showpacs]{revtex4}
\usepackage{epsfig}

\begin{document}
\draft
\title{Charge sensitivity of the Inductive Single-Electron Transistor}

\author{Mika A. Sillanp\"a\"a}
\author{Leif Roschier}
\author{Pertti J. Hakonen}
\affiliation{Low Temperature Laboratory, Helsinki University of
Technology, Otakaari 3 A, Espoo P.O.Box 2200 FIN-02015 HUT
Finland}

\begin{abstract}
We calculate the charge sensitivity of a recently demonstrated
device where the Josephson inductance of a single Cooper-pair
transistor is measured. We find that the intrinsic limit to
detector performance is set by oscillator quantum noise.
Sensitivity better than $10^{-6}$e$/\sqrt{\mathrm{Hz}}$ is
possible with a high $Q$-value $\sim 10^3$, or using a SQUID
amplifier. The model is compared to experiment, where charge
sensitivity $3 \times 10^{-5}$e$/\sqrt{\mathrm{Hz}}$ and bandwidth
100 MHz are achieved.
\end{abstract}


\maketitle


Remarkable quantum operations have been demonstrated in the solid
state \cite{nakamuraqb,vion,hanqb}. As exotic quantum measurements
known in quantum optics are becoming adopted for electronic
circuits \cite{qed}, sensitive and desirably non-destructive
measurement of the electric charge is becoming even more
important.

A new type of fast electrometer, the Inductive Single-Electron
Transistor (L-SET) was demonstrated recently \cite{lset}. Its
operation is based on gate charge dependence of the Josephson
inductance of a single Cooper-pair transistor (SCPT). As compared
to the famous rf-SET \cite{rfset}, where a high-frequency
electrometer is built using the control of single-electron
dissipation, the L-SET has several orders of magnitude lower
dissipation due to the lack of shot noise, and hence also
potentially lower back-action.

Charge sensitivity of the sequential tunneling SET has been
thoroughly analyzed. However, little attention has been paid on
detector performance of the SCPT, probably because no real
electrometer based on SCPT had been demonstrated until invention
of the L-SET. Some claims have been presented \cite{lset,moriond}
that performance of SCPT in the L-SET setup could exceed the
shot-noise limit of the rf-SET \cite{paalanen}, $s_q \simeq
10^{-6}$e$/\sqrt{\mathrm{Hz}}$, but no accurate calculations have
appeared.

In this letter we carry out a sensitivity analysis for L-SET in
the regime of linear response. We find that (neglecting $1/f$
background charge noise) the intrinsic limit to detector
sensitivity is set, unlike by shot noise of electron tunneling in
a normal SET, by zero-point fluctuations \cite{zorin}.

\begin{figure}[h]
  \includegraphics[width=8.5cm]{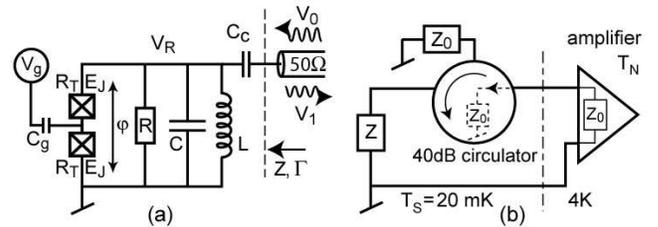}
  \caption{The L-SET resonator (a), and its equivalent circuit
(impedance $Z$) coupled to cabling (b).}\label{fig:schema}
\end{figure}

A SCPT has the single-junction Josephson energy $E_J$, and the
total charging energy $E_C = e^2/(2 C_{\Sigma})$, where
$C_{\Sigma}$ is the total capacitance of the island. At the lowest
energy band the energy is $E_0$, the effective Josephson energy is
$E_J^{*} =
\partial ^2 E_0(q, \varphi) / \partial \varphi^2$, and the effective
Josephson inductance is $L_J = (2 \pi / \Phi_0)^2 (E_J^{*})^{-1}$.
These have a substantial dependence on the (reduced) gate charge
$q = C_g V_g / e$ if $E_J / E_C \lesssim 1$. Here, $\varphi$ is
the phase across the SCPT. With a shunting capacitance $C$, SCPT
forms a parallel oscillator. We further shunt the oscillator,
mainly for practical convenience, by an inductor $L \simeq L_J$.
Hence we have the resonator as shown in Fig.~\ref{fig:schema},
with the plasma frequency $f_p = \omega_p /(2\pi) = 1/(2 \pi)
(L_{\mathrm{tot}} C)^{-1/2} \approx 1$ GHz, where
$L_{\mathrm{tot}} = L
\parallel L_{J}$.

The coupling capacitor, typically $C_c \ll C$, allows, in
principle, for an arbitrarily high loaded quality factor $Q_L$. If
directly coupled to feedline, $Q_L = Z_0 \sqrt{C/L_{\mathrm{tot}}}
\approx 1$, which is clearly intolerable. With a coupling
capacitor, however, $Q_L = 1/2 Q_i$ in the optimal case (as shown
later) of critical coupling $Z = Z_0$ ($Q_i$ is the internal
$Q$-value). The resistor $R$ is a model component for internal
losses.

We consider only the linear regime, where the detector works by
converting charge to resonant frequency. We do not model the
"anharmonic" operation mode \cite{lset} where $\varphi$ oscillates
in the nonlinear regime of the Josephson potential, yielding in
fact better sensitivities in experiment.

Impedance of the L-SET circuit as illustrated in
Fig.~\ref{fig:schema} is

\begin{equation}\label{eq:Z}
    Z = \frac{1}{i \omega C_c} + \left( i \omega C +
    \frac{1}{i \omega L} + \frac{1}{i \omega L_{J}} + \frac{1}{R} \right)^{-1}.
\end{equation}

\noindent The circuit is probed by measuring the voltage
reflection coefficient $\Gamma = |\Gamma |\exp(i \arg (\Gamma))=(Z
- Z_0) / (Z + Z_0)$ to an incoming voltage wave of amplitude
$V_0$. The reflected wave amplitude is $V_1 = |\Gamma| V_0$. Here,
$Z_0 = 50 \, \Omega$ is the wave impedance of coaxial lines.

Spectral density of noise power at the output of the 1st stage
amplifier, referred to amplifier input, is $k_B T_N^*$, where the
effective temperature $T_N^*$ is due to amplifier noise and sample
noise: $T_N^* = T_N + T_S$. Sample is supposed to be critically
coupled, and hence its noise is like that of a $50 \, \Omega$
resistor at the temperature $T_S$ (note that the Josephson effect
is a system ground state property and hence it contributes no
noise). Typically, $k_B T_S \lesssim \hbar \omega_p$, and thus
sample noise is already in the quantum limit.

Noise of contemporary rf-amplifiers, however, remains far from the
quantum limit, i.e., $T_N \gg T_S$. The best demonstrated
SQUID-based rf-amplifiers have reached $T_N \sim 100-200$ mK
\cite{squid}. Therefore, added noise from the sample can be safely
ignored when analyzing detector performance.

Charge sensitivity for \emph{amplitude modulation} (AM) of the
rf-SET was calculated in detail in Ref.\ \cite{leif} assuming
detection of one sideband. It was assumed that the sensitivity is
limited by general equivalent noise temperature similarly as here,
and hence the formula applies as such:

\begin{equation} \label{eq:sq}
  s_q = \sqrt{2 k_B T_N} / \left(V_0 \frac{\partial | \Gamma|}{\partial q} \right).
\end{equation}

\noindent In the linear regime, the best sensitivity of the L-SET
is clearly at the largest acceptable value of $V_0$, where
linearity still holds reasonably well. This is the case when an AC
current of critical current peak value flows through the SCPT, and
the phase swing is $\pi$ p-p. Then, voltage across the SCPT, and
resonator (later we discuss important quantum corrections to this
expression), $V_R = \left| 2 Z_R /(Z + Z_0) \right| V_0$ equals a
universal critical voltage of a Josephson junction
\cite{MSthesis}, $V_C = \pi \hbar \omega / (4 e) \simeq 3 \, \mu$V
at $f_p = 1$ GHz. Here, $Z_R$ is impedance of the parallel
resonator.

We decompose the derivative in Eq.~(\ref{eq:sq}) into terms due to
the circuit and SCPT: $\frac{\partial | \Gamma|}{\partial q} =
\frac{\partial | \Gamma|}{\partial \omega_p} \frac{\partial
\omega_p}{\partial L_J}\frac{\partial L_J}{\partial q}$. We define
a dimensionless transfer function $g' = (\partial L_{J} / \partial
q) (e / L_{J0})$ scaled according to minimum (w.r.t. gate) of
$L_J$. The gate value which yields the maximum of $g'$, denoted
$g$, is the optimum gate DC operation point of the charge
detector. In what follows, $L_J$ should be understood as its value
at this point. With a given $E_J / E_C$ ratio, we compute the
values of $g$ and $L_J$ numerically from the SCPT band structure
($g$ is plotted in Fig.\ 4 in Ref.\ \cite{lset}). If $E_J /E_C \ll
1$, one can use the analytical result $L_{J0} = 2 \Phi_0^2 /(\pi
E_J)$.

With a general choice of parameters of the tank resonator,
Eq.~(\ref{eq:sq}) needs to be evaluated numerically. However, when
the system is critically coupled, $Z = Z_0$, a simple analytical
formula can be derived. Numerical calculations of
Eq.~(\ref{eq:sq}) over a large range of parameters show that the
best sensitivity occurs when $Z = Z_0$. This is reasonable because
it corresponds to the best power transfer. All the following
results are for critical coupling. Later, we examine effects of
detuning from the optimum. Initially, we also suppose the
oscillator is classical, i.e., its energy $E \gg \hbar \omega_p$.

Optimal value of the coupling capacitor is calculated using $Q_L =
1/2 Q_i$, and we get $C_c = \sqrt{C / (\omega_p Q_i Z_0)}$.

Since it was assumed $Z = Z_0$, it holds that $Z_R = Z_0 +
i/(\omega_p C_c)$. Voltage amplification by the resonator then
becomes $V_R = V_0 \sqrt{Q_i / (\omega_p Z_0 C)}$ which holds for
a reasonably large $Q_i$. We thus have $V_0 = \pi \hbar
\omega_p^{3/2} \sqrt{Z_0 C} / (4 e \sqrt{Q_i})$.

With $\omega_p = (L_{\mathrm{tot}} C)^{-1/2}$, we get immediately
$(\partial \omega_p / \partial L_J)^{-1} = 2 \sqrt{C} L_J^2
\sqrt{1/L + 1/L_J}$. Using the fact \cite{qfactor} that FWHM of
the loaded resonance absorption dip at critical coupling is
$\omega_p / (2 Q_L)$, we get $\partial | \Gamma| /
\partial \omega_p = 2 Q_L / \omega_{p} = Q_i / \omega_{p}$.

Inserting these results into Eq.~(\ref{eq:sq}), we get expression
for the AM charge sensitivity in the limit the oscillator is
classical:

\begin{equation}\label{eq:sq1}
    s_q = \frac{8 e L_J^2 \sqrt{\frac{1}{L} +
    \frac{1}{L_J}} \sqrt{2 k_B T_N} }
    {g \pi \hbar L_{J0} \sqrt{\omega_p Q_i}}
\end{equation}

\noindent in units of $[e / \sqrt{\rm Hz}]$. Clearly, the shunting
inductor is best omitted, i.e., $L \rightarrow \infty$. The
classical result, Eq.~(\ref{eq:sq1}), improves without limit at
low $E_J/E_C$.

We will now discuss quantum corrections to Eq.~(\ref{eq:sq1}).
Although \emph{spectral density} of noise in the resonator is
negligible in output, \emph{integrated} phase fluctuations even
due to quantum noise can be large. Integrated phase noise in a
high-$Q$ oscillator is $\langle \Delta \varphi ^2 \rangle = 2
\pi^2 \hbar L_{\mathrm{tot}} \omega_p / \Phi_0^2$ \cite{devoret}.
When $\langle \Delta \varphi \rangle$ exceeds the linear regime
$\sim \pi$, which happens at high inductance (low $E_J / E_C$),
plasma resonance "switches" into nonlinear regime, and the gain
due to frequency modulation vanishes. If $L \gg L_J$, and $f_p
\sim 1$ GHz, we have ultimate limits of roughly $E_J / E_C \sim
0.06$, or $\sim 0.02$, for a SCPT made out of Al or Nb,
respectively.

Even before this switching happens, quantum noise in the
oscillator $E_Q = \frac{1}{2} \hbar \omega_p$ has an adverse
effect because less energy can be supplied in the form of drive,
that is, $V_0$ is smaller. This can be calculated in a
semiclassical way as follows. Energy of the oscillator is due to
drive ($E_D$) and noise (we stay in the linear regime): $E =
(\Phi_0 \varphi)^2 / (8 \pi^2 L_{\mathrm{tot}}) = E_D + E_Q
=(\Phi_0 \varphi_D)^2 / (8 \pi^2 L_{\mathrm{tot}}) + \frac{1}{2}
\hbar \omega_p$, where the phases are in RMS, $\varphi$ is the
total phase swing, and $\varphi_D$ is that due to drive. Solving
for the latter, we get $\varphi_D = \sqrt{\varphi^2 - 4 \pi^2
\hbar \omega_p L_{\mathrm{tot}} / \Phi_0^2}$. The optimal drive
strength $V_R = V_C$ corresponds to $\sqrt{2} \varphi = \pi/2$,
and hence the maximum probing voltage $V_0$ is reduced by a factor
$\beta = \sqrt{1- 32\hbar \omega_p L_{\mathrm{tot}} / \Phi_0^2}$
due to quantum noise in the oscillator.

\begin{figure}[h]
  \includegraphics[width=8.5cm]{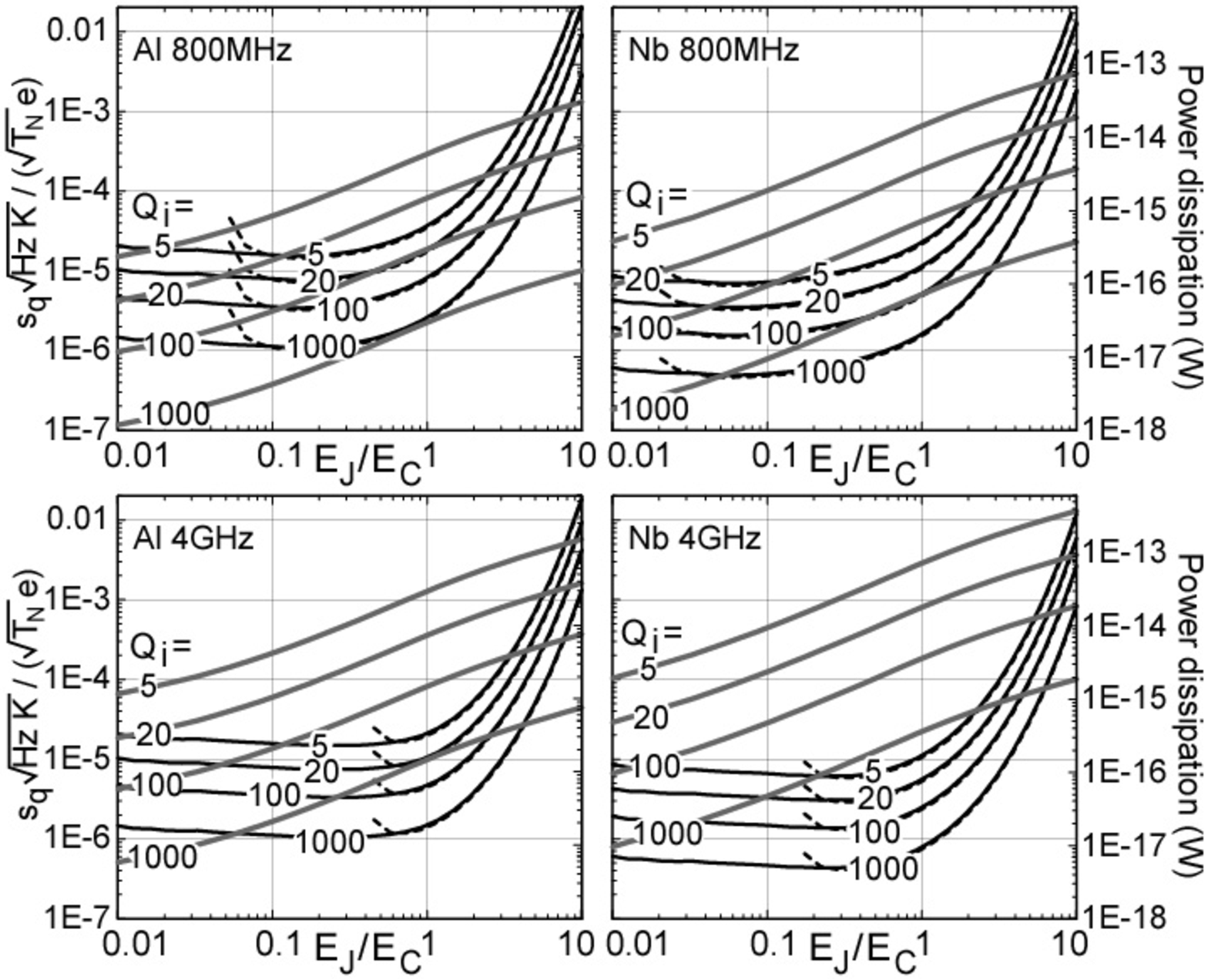}
  \caption{Charge sensitivity of the L-SET optimized from Eq.~(\ref{eq:sq}) (black lines).
  The analytical result (Eq.~(\ref{eq:sq1}) multiplied by $\beta ^{-1}$), with $L = \infty$,
  is shown with dashed lines. Gray lines are the corresponding
  power dissipation. All the graphs have the same scales, which are
indicated for $s_q$ (left) and dissipation (right).
  The curves are for different $Q_i$ as marked. All graphs have $Z \simeq Z_0$.}
\label{fig:sqpredicted}
\end{figure}

The optimal sensitivity is finally

\begin{equation}\label{eq:sqFUND}
    s_q^{QL} = \frac{64 \sqrt{2} e L_J^2 \sqrt{2 k_B T_N} }
    {g \pi \sqrt{\hbar} \Phi_0 L_{J0} \sqrt{Q_i}},
\end{equation}

\noindent which depends only weakly on operation frequency. We
optimized Eq.~(\ref{eq:sq}) (replacing there $V_0$ by $\beta V_0$)
assuming similar tunnel junction properties as in the experiment,
$E_J E_C = 1.8$ K$^2$ (Al) and $E_J E_C = 10$ K$^2$ (Nb). The
results are plotted in Fig.~\ref{fig:sqpredicted} together with
corresponding power dissipation $(V_C / \sqrt{2})^2 / R = \pi^2
\hbar^2 \omega_p /(32 e^2 Q_i L_J)$.

\begin{figure}[h]
  \includegraphics[width=8.5cm]{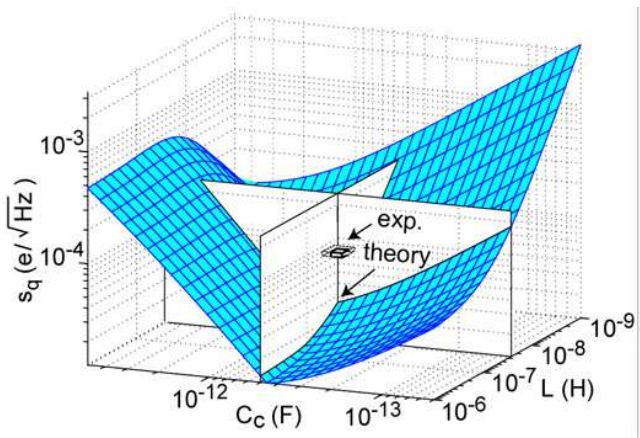}
  \caption{Measured charge sensitivity (box) compared
to calculations. In experiment, $C_c \sim 0.5$fF, and $L \sim
28$nH. The graph shows also how the sensitivity would change as a
function $C_c$ and $L$ according to the model.}
\label{fig:sqillust}
\end{figure}

The optimal sensitivity is reached around $E_J / E_C \simeq 0.1
... 0.3$, where the curves in Fig.~\ref{fig:sqpredicted} almost
coincide Eq.~(\ref{eq:sqFUND}). $C_c$ should be chosen so that
critical coupling results. Typically also it should hold $L \gg
L_J$ (see the analytical curve in Fig.~\ref{fig:sqpredicted}).
However, sensitivity decreases only weakly if these values are
detuned from their optimum (Fig.~\ref{fig:sqillust}).

By numerical investigation we found that readout of $\arg
(\Gamma)$, with mixer detection, offers within accuracy of
numerics the same numbers than the discussed AM (readout of
$|\Gamma|$).

In experiment, we measured charge sensitivity for the following
sample and resonator: $R_T \simeq 11$ k$\Omega$, $E_J \simeq 0.7$
K, $E_C \simeq 2.6$ K, $E_J / E_C \simeq 0.3$, $Q_i \simeq 16$, $L
\simeq 28$ nH, $C \simeq 1.2$ pF, $C_c \simeq 0.5$ pF. In all
samples so far, $Q_i \lesssim 20$ which is presently not
understood. The measurements were done as described in Ref.\
\cite{lset}, with $T_N \sim 5$ K \cite{amplifier}. We measured
$s_q = 7 \times 10^{-5}$e$/\sqrt{\mathrm{Hz}}$ by AM at 1 MHz,
while a prediction with the present parameters is $s_q = 3 \times
10^{-5}$e$/\sqrt{\mathrm{Hz}}$ (see also Fig.~\ref{fig:sqillust}).

Theory and experiment thus agree reasonably. The somewhat lower
sensitivity in experiment is likely to be due to external noise
which forces a lower $V_0$ and also smoothes out the steepest
modulation. Its origin is not clear. Also the $25 \%$ higher
values of $L_J$ than expected agree qualitatively with noise.

In the "anharmonic" mode, we measured $s_q = 3 \times
10^{-5}$e$/\sqrt{\mathrm{Hz}}$, with a usable bandwidth of about
100 MHz ($s_q \sim 10^{-4}$e$/\sqrt{\mathrm{Hz}}$ at 100 MHz).
Considering both $s_q$ and band, a performance comparable to the
best rf-SETs \cite{rfset,chalmers} has been reached with the
L-SET, though here at more than two orders of magnitude lower
power dissipation ($\sim 10$ fW).

In the linear regime, the power lost $P_{\Sigma}$ from drive
frequency $m = 1$ to higher harmonics is determined by the sum,
for $m \geq 2$, of Josephson junction admittance components $ |
Y_m | = 2 J_m \left(2 e /(\hbar \omega) V_{1} \right)$. At the
critical voltage $V_{1} = V_C$, this amounts to $Y_{\Sigma} / Y_1
= P_{\Sigma} / P_1 \sim 30$ \%. Since charge sensitivity is
proportional to square root of power, it thus decreases only $\sim
15$ \% due to non-linearity. Further corrections due to slightly
non-sinusoidal lowest band of the SCPT, as well as asymmetry due
to manufacturing spread in junction resistance, we estimate as
insignificant.

Next we discuss non-adiabaticity. Interband Zener transitions
might make the SCPT jump off from the supposed ground band 0. We
make a worst case estimate by assuming that the drive is $2 \pi$
p-p (partially due to noise). The probability to cross the minimum
$\Delta_m$ of band gap $\Delta = E_1 - E_0$ is: $P_Z \simeq \exp
\left( - \pi \Delta_m^2 / (2 \hbar D \dot{\varphi}) \right)$,
where we evaluate the dependence of the band gap on phase $D =
\partial \Delta / \partial \varphi$ at $\varphi = \pi/2$.
$\dot{\varphi} = 2 \omega_p$ is determined by the drive.

Zener tunneling is significant if it occurs sufficiently often in
comparison to $1 \rightarrow 0$ relaxation. Threshold is when $P_Z
\sim \Gamma_{\downarrow} / (2 f_0)$, where $\Gamma_{\downarrow}
\gtrsim (1 \mu \rm{s})^{-1}$ is the relaxation rate. Operation of
the L-SET can thus be affected above $P_Z \sim 10^{-4}$.

Numerical calculations for $P_Z$ show that Zener tunneling is
exponentially suppressed, at the L-SET optimal working point, in
the interesting case of low $E_J/E_C$ \cite{MSthesis}. This is
because $\Delta_m$ becomes large and $D$ small. For instance, if
$E_J = 1$ K and $f_p = 1$ GHz, we got that Zener tunneling is
insignificant below $E_J/E_C \sim 3$. With $E_J = 0.5$ K and $f_p
= 5$ GHz, the threshold is $E_J/E_C \simeq 1$.

We conclude that with sufficiently high $Q_i$ and using a
amplifier close to the quantum limit, even $s_q \sim
10^{-7}$e$/\sqrt{\mathrm{Hz}}$, order of magnitude better than the
shot-noise limit of rf-SET, is intrinsically possible for the
L-SET. So far, the sensitivity has been limited by $Q_i \lesssim
20$.

Fruitful discussions with M. Feigel'man, U. Gavish, T. Heikkil\"a,
R. Lindell, H. Sepp\"a are gratefully acknowledged. This work was
supported by the Academy of Finland and by the Large Scale
Installation Program ULTI-3 of the European Union (contract
HPRI-1999-CT-00050).

\end{document}